# Nitrogen oxides under pressure: stability, ionization, polymerization, and superconductivity

Dongxu Li[1,*], Artem R. Oganov[2,3,4,5,6], Xiao Dong[7], Xiang-Feng Zhou[3,4,7], Qiang Zhu[3,4], Guangrui Qian[3,4], Huafeng Dong[3,4]

[1] College of Materials Science and Engineering, Huaqiao University, Xiamen, 361021 P.R. China

[2] Skolkovo Institute of Science and Technology, Skolkovo Innovation Center, 3 Nobel St., Moscow 143026, Russia

[3] Department of Geosciences, Stony Brook University, Stony Brook, NY 11794, USA

[4] Center for Materials by Design, Institute for Advanced Computational Science, Stony Brook University, Stony Brook, NY 11794, USA

[5] Moscow Institute of Physics and Technology, 9 Institutskiy lane, Dolgoprudny city, Moscow Region, 141700, Russia

[6] School of Materials Science and Engineering, Northwestern Polytechnical University, Xi'an, 710072, China

[7] School of Physics and Key Laboratory of Weak-Light Nonlinear Photonics, Nankai University, Tianjin 300071, China

[*]Corresponding author: lidongxu@hqu.edu.cn

**ABSTRACT**: Nitrogen oxides are textbook class of molecular compounds, with extensive industrial applications. Nitrogen and oxygen are also among the most abundant elements in the universe. We explore the N-O system at 0 K and up to 500 GPa though ab initio evolutionary simulations. Results show that two phase transformations of stable molecular $NO_2$ exist at 7 and 64 GPa, and followed by decomposition of $NO_2$ at 91 GPa. All of the $NO^+NO_3^-$ structures are found to be metastable at T=0 K, so experimentally reported ionic $NO^+NO_3^-$ is either metastable or stabilized by temperature. Upon increasing pressure, $N_2O_5$ transforms from *P*-1 to *C*2/*c* structure at 51 GPa. NO becomes thermodynamically stable at 198 GPa. This polymeric phase is superconducting ($T_c = 2.0$ K) and contains a -N-N- backbone.

Both nitrogen and oxygen have been extensively investigated in experiments and theoretical simulations. Generally, nitrogen is an insulator or a semiconductor. Cubic gauche phase of nitrogen [1] is stable in a wide range of pressure [2]. Other nitrogen structures, such as chain and rings[3-5], have also been reported. All known phases of oxygen are molecular [6,7]. Experiments and first-principles calculations for oxygen under high pressure revealed the complex evolution of insulator-semiconductor-metal-semiconductor [8]. The superconductivity of solid oxygen ($T_c = 0.6$ K) was observed at above 96 GPa in experiment [9]. The known nitrogen oxides are semiconducting (for example, the band gap of *Im*-3 $NO_2$ calculated is approximately 2.8 eV).

At ambient pressure, nitrogen oxides exist as molecular crystals with many applications in chemical industry and important biological roles. The volumetric behavior of nitrous oxide under pressure has been investigated since 1961 [10]. The synthesis and phase transformations of $N_2O$ have been analyzed in experimental and theoretical studies[11-16]. Different from normal phases containing $N_2O_4$ molecules, the ionic $NO^+NO_3^-$ was reported in the range of 1.5 to 3.0 GPa [17]. The typical N-O stretch of NO$^+$ was characterized at 2234 cm$^{-1}$, consistent with previous reports[17,18]. In 2001, Somayazulu et al. [19] synthesized the ionic NO$^+$NO$_3^-$ (nitrosonium nitrate) phase from N$_2$O at above 20 GPa and 1000 K, and performed the first structural characterization of NO$^+$NO$_3^-$. Somayazulu et al. [19] proposed an ionic NO$^+$NO$_3^-$ model based on aragonite with space group of $P2_1cn$. Other $P2_1/m$ [20] and $Pna2_1$ [16] models of NO$^+$NO$_3^-$ were also suggested, and the later one is more stable. However, the simulated XRD data of the $Pna2_1$ structure is quite different from that in experiments, indicating that other undiscovered stable NO$^+$NO$_3^-$ structures might exist.

**Results and Discussions**

We employed the evolutionary algorithm USPEX[21-24] to predict the stable N-O compounds and structures under high pressures. Up to 500 GPa, only three stable N-O compounds were found (NO$_2$, N$_2$O$_5$ and NO), as seen in Fig. 1. Most of them retain their molecular structures even under high pressure. Experimentally known "laughing gas" N$_2$O is metastable. The stable phases are discussed as follows.

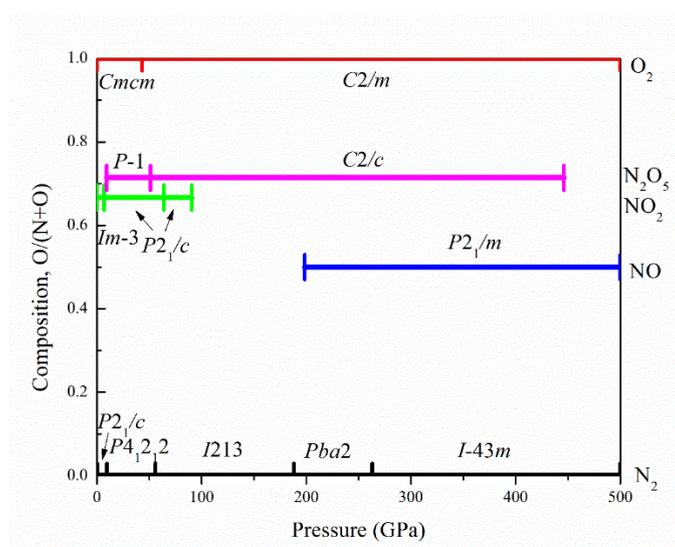

Figure 1 Phase diagram of the N–O system.

(1) NO$_2$: Besides the known cubic (*Im*-3) and monoclinic (*P*2$_1$/*c*) NO$_2$ structures are stable in pressure ranges of 0-7 and 7-64 GPa respectively, another *P*2$_1$/*c* structure was found to be stable from 64 to 91 GPa (Fig. 2a). Similar to the known phases, this novel NO$_2$ structure also contains N$_2$O$_4$ molecules. Different from known *P*2$_1$/*c* NO$_2$, the proposed *P*2$_1$/*c* NO$_2$ is denser and has 8 formula units in the unit cell. NO$_2$ decomposes at 91 GPa.

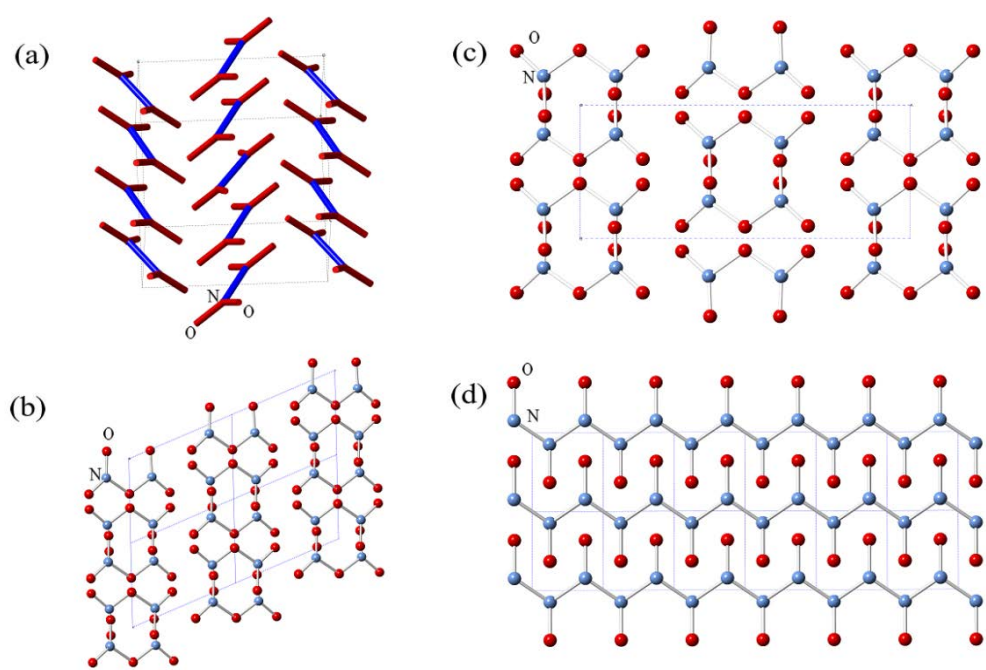

Figure 2 Structures of (a) *P*2$_1$/*c* NO$_2$, (b) *P*-1 and (c) *C*2/*c* N$_2$O$_5$, (d) *P*2$_1$/*m* NO.

(2) N$_2$O$_5$: Molecular N$_2$O$_5$ phases are stable in a wide pressure range (9-446 GPa). At 51 GPa, N$_2$O$_5$ transforms from *P*-1 (Fig. 2b) to *C*2/*c* (Fig. 2c) structure. The N$_2$O$_5$ molecules remain planar. At 446 GPa, N$_2$O$_5$ becomes unstable and decomposes into NO and O. At 0 GPa, the *P*-1 N$_2$O$_5$ is 0.04 eV/atom more stable than known hexagonal NO$_2$NO$_3$, however, both of them are calculated to be above the convex hull, and therefore metastable.

(3) NO: NO is a metastable compound at low pressures. A polymeric NO structure

(Fig. 2d, $P2_1/m$) becomes stable at 198 GPa. Nitrogen atoms form a strong covalent backbone (N-N=1.34 Å) in the shape of a zigzag chain. Indeed, that is right between the typical values of single (1.45 Å) and double (1.25 Å) nitrogen-nitrogen bonds. Each nitrogen atom is also bonded to one oxygen atom (N-O bond length is 1.20 Å). A similar backbone has also been reported for the N-H system [25]. Distance between neighboring quasi-one-dimensional structures is 1.86 Å. The phonon dispersion curve of this remarkable polymeric phase was calculated (shown in Fig. S6). No imaginary frequency was observed, implying its dynamical stability.

While most of the stable N-O phases are semiconducting, polymeric NO is metallic. The band structure of NO is shown in Fig. 3. Using the Allen-Dynes modified McMillan equation[26,27] with value of the Coulomb pseudopotential with $\mu^*$=0.13, polymeric NO is superconducting with $T_c$ =2.0 K at 200 GPa, which is close to that of oxygen [8,9].

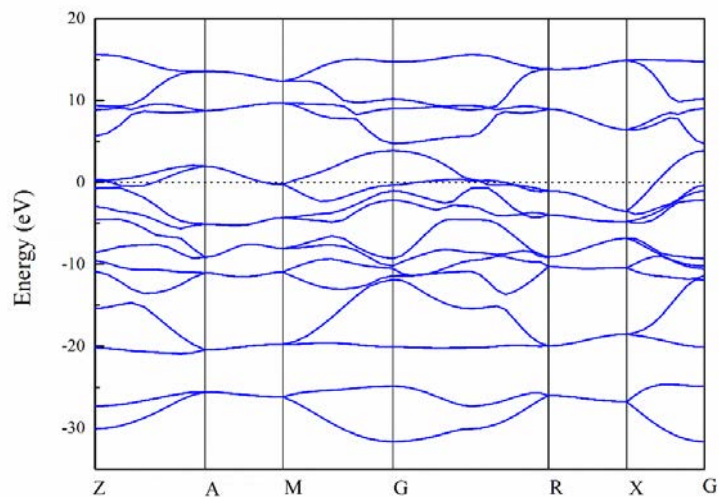

Figure 3 Band structure of $P2_1/m$ NO at 198 GPa. Z(0,0,0.5), A(0.5,0.5,0.5), M(0.5,0.5,0), G(0,0,0), R(0,0.5,0.5) and X(0,0.5,0).

As mentioned above, ionic $NO^+NO_3^-$ has been observed in several high pressure experiments[19,20,28]. However, no stable $NO^+NO_3^-$ structure was found in our variable-composition searches. To find the lowest-enthalpy ionic $NO^+NO_3^-$ structure, we performed $(NO)_n(NO_3)_n$ (n=6 or 8) calculations at 0-50 GPa, assembling structures from ready-made NO and $NO_3$ units in variable proportion. A novel metastable monoclinic $NO^+NO_3^-$ ($P2_1$, Fig. 4) was found to be more stable than orthorhombic phase [16] and monoclinic $P2_1/m$ $NO^+NO_3^-$ [20] at pressures above 1.7 GPa. The main difference between novel $P2_1$ and $P2_1/m$ $NO^+NO_3^-$ models [20] is the orientation of the $NO^+$ molecules. Importantly, at all pressures structures made of $N_2O_4$ molecules are more stable than ionic $NO^+NO_3^-$ structures (Fig.4)

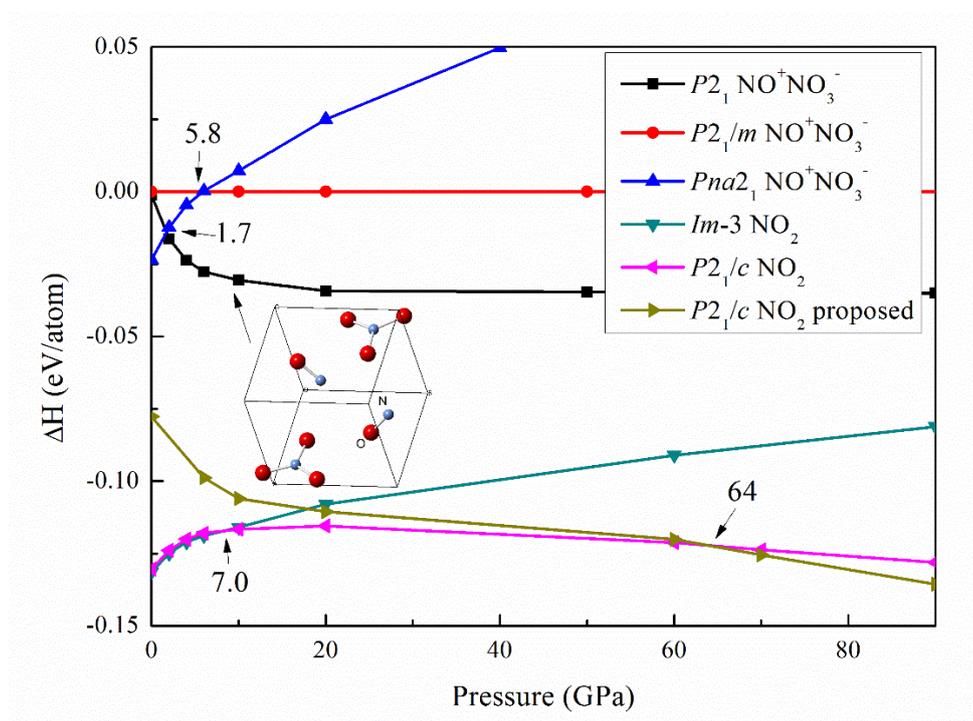

Figure 4 Enthalpies of $NO_2$ phases as a function of pressures.

In experiments, the typical Raman frequencies of $NO^+NO_3^-$ are 2234 cm$^{-1}$ for the N-O stretch in $NO^+$, together with 1345, 1056 and 721 cm$^{-1}$ for anti-symmetric stretch, symmetric stretch and in-plane deformation for $NO_3^-$ respectively[17,19,28-30]. The Raman frequencies and intensities of $NO^+NO_3^-$ and $NO_2$ structures were calculated at 20 GPa. Here, Raman frequencies of $NO^+$ and $NO_3^-$ were used for comparison. As shown in Fig. 5, the simulation data from of $P2_1/c$ $NO_2$ and $Pna2_1$ $NO^+NO_3^-$ are significantly different from that in the experiments [17]. The typical Raman frequencies of N-O stretch are 2071 cm$^{-1}$ of $P2_1$ and 2151 cm$^{-1}$ of $P2_1/m$ structures. Both of them basically match the experimental data[19], but that of $P2_1$ $NO^+NO_3^-$ obtains better match in terms of relative intensity. Similar comparison results for Raman and XRD data could also be observed under other pressures[20,31].

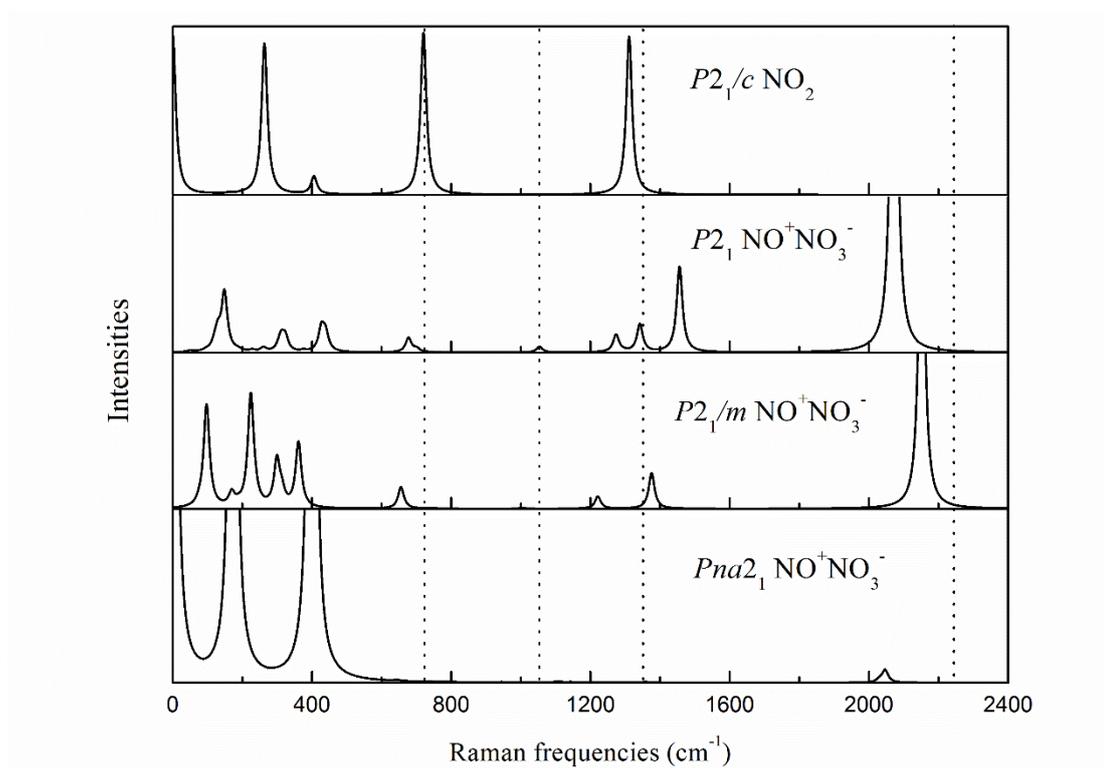

Figure 5 Simulated Raman spectra of $P2_1/c$ $NO_2$, and $P2_1$, $P2_1/m$ [20] and $Pna2_1$ [16] $NO^+NO_3^-$ at 20 GPa. Typical Raman frequencies of $NO^+$ and $NO_3^-$ in experiment [17] were drawn by dotted lines.

In summary, stable $NO_2$, $N_2O_5$ and NO phases were found in N-O system up to

500 GPa. The $P2_1/c$ NO$_2$ becomes stable at 64 GPa and decomposes at 91 GPa. N$_2$O$_5$ with $P$-1 becomes stable at 9 GPa, transforms to $C2/c$ at 51 GPa and decomposes at 446 GPa. The only metallic structure ($P2_1/m$ NO) has -N-N- zigzag backbone and possesses superconductivity with $T_c$ = 2.0 K. Our results show that ionic NO$^+$NO$_3^-$ is metastable, and we identify a novel $P2_1$ structure that matches experimental data better and has lower enthalpy than previously proposed structures.

**Methods**

An evolutionary algorithm, as implemented in the USPEX code[21-24], were performed to search for the stable compounds and structures. This method has already been successfully applied to study numerous systems, including nitrogen and oxygen under pressure [5,7]. Structure relaxations were done using density functional theory (DFT) [32,33] within the generalized gradient approximation (GGA) [34] using the all-electron projector augmented wave (PAW) [35,36] method as implemented in the VASP code [37]. The plane-wave kinetic energy cutoff was set to 600 eV and Brillouin zone was sampled at a resolution of 2π×0.06 Å$^{-1}$. At first, variable-composition were carried out at 0, 10, 20, 30, 50, 100, 150, 200, 250, 300, 350, 400, 450 and 500 GPa. Stability of compounds was judged using the convex hull construction: those compounds which are on the convex hull (i.e. which are more favorable than any isochemical mixture of other phases) are thermodynamically stable at given conditions. The PHONOPY code [38] was employed to calculate phonon dispersions for all promising structures, and all the discussed structures were found to be dynamically stable. All Raman frequencies and intensities were calculated according to the method of Porezag and Pederson[39].

The electron–phonon coupling calculations in Quantum Espresso[40] with 180 Ry plane-wave cutoff energy were used to calculate the critical temperature of superconductivity ($T_c$).

**Acknowledgement**

The authors appreciate financial supports from Natural science foundation of Fujian Province (No. 2013J05075) and Fujian Education Department (No. JA13016). The authors would like to thank Prof. Jincao Dai and Qianku Hu for their helpful suggestions.

**Author contributions**

D.X.L., X.D., G.R.Q., Q.Z., H.F.D. and X.F.Z. performed calculations and analyzed the data. D.X.L., A.R.O. and Q.Z. wrote the paper. All authors reviewed the manuscript.

**Additional Information**

Supplementary information accompanies this paper: Convex hull diagram of N-O system, lattice parameters and phonon dispersion curves of $NO_2$, $N_2O_5$, NO and $NO^+NO_3^-$, and X-Ray Diffraction data of $NO_2$.

Competing financial interests: The authors declare no competing financial interests.